\documentclass[a4paper, conference]{IEEEtran}
\usepackage{amsmath}
\usepackage{amssymb}
\usepackage{mathdots}
\usepackage{graphicx}
\usepackage{cite}
\usepackage{subfig}
\usepackage{cuted}
\usepackage{algorithm}
\usepackage{algpseudocode}
\begin{document}
%%%%%%%%%%%%%%%%%%%%%%%%%%%%%%%%%%%%%
\title{On the Sensitivity of Active RIS Systems to CSI Errors: Joint Optimization and Performance Trade-off}

\author{\IEEEauthorblockN{ Mohamed Shalma, \  Engy Aly Maher, and \ Ahmed El-Mahdy}
\IEEEauthorblockA{\textit{Faculty of Information Engineering and Technology (IET)} \\
\textit{German university in Cairo}\\
\textit{Cairo, Egypt}\\
\{mohamed.hamed;engy.aly;ahmed.elmahdy\}@guc.edu.eg}

}

\maketitle

\begin{abstract}
In this paper, the problem of maximizing the sum-rate is addressed for a multi-user uplink scenario that is assisted by an active reconfigurable intelligent surface (RIS). The maximization is achieved by optimizing the beamforming at the base station, the users' transmit power, active RIS elements phase shifts, and active gains in presence of imperfect channel state information (CSI). The non-convex maximization problem is decomposed into sub-problems and solved via iterative approaches including the Lagrangian method, the projected gradient descent, multi-variate Taylor expansion and fractional programming. Numerical results show that the active RIS is more sensitive to CSI imperfections than passive one at high error variances.
\end{abstract}


\IEEEpeerreviewmaketitle

\section{Introduction}

Tremendous attention has been recently given to reconfigurable intelligent surfaces (RIS) as a promising technology to control the communication channel and enhance the wireless network performance. RIS is classified into two types, active and passive \cite{9903846,9928771}. For active RIS, the reflecting elements (RE) can amplify the signal falling on the RIS besides introducing phase shifts to it. This property is desirable to overcome the double fading effect of the RIS and strengthen the received signal power. Consequently, active RISs received higher attention than passive ones \cite{10525785,11313425,10525765}

However, obtaining perfect channel state information (CSI) is very challenging. Furthermore, passive and active RISs can have different channel estimation techniques.
both \cite{10959043,10025392} discussed active RIS scenario with phase noise and imperfect CSI based on linear minimum mean square error (LMMSE). The authors focus on statistical analysis rather than optimization. Furthermore, \cite{10025392} is limited to single-user scenario. More importantly, \cite{10959043} assumes LMMSE estimation of the aggregated channel (direct plus RIS-assisted links). However, in practical systems, channel estimation is typically performed in multiple stages \cite{9722893,9500469}, as discussed in Section II. For effective joint resource optimization, the direct channel, RIS-assisted channels, and RIS–BS channel are generally required individually, since relying solely on the aggregated channel limits the flexibility of optimizing the system parameters. Furthermore, as shown later, the CSI errors become inherently coupled with the power allocation, beamforming design, and RIS element configuration under the practical three-stage estimation strategy, which further complicates the joint optimization problem. 
In \cite{10742558}, the authors derived approximate expressions for users' outage probability and ergodic capacity for active RIS scenario with imperfect CSI under Nakagami fading channel.

meanwhile \cite{10757678,10992287} proposed optimization frameworks for active RIS-aided scenarios with imperfect CSI, the main focus is shifted towards the deep reinforcement learning. The authors in \cite{10757678,10992287} focus on DRL validity and enhancement in vehicular communication through active RIS rather than focusing on trade-off of imperfect CSI effect regarding active and passive RISs. In this work, we focus primarily on performance gain of CSI error for active and passive RISs with coupled CSI error among resources and the sensitivity to it.

All previous works \cite{10959043,10025392,10742558} focus on statistical analysis rather than optimization of active RIS-aided system with imperfect CSI. To the best of our knowledge, the joint optimization of power, beamforming vectors, the gain and phase shifts to maximize the sum-rate of an uplink active RIS-assisted scenario with imperfect CSI knowledge has not been fully addressed. The contributions of this paper are:
\begin{itemize}
    \item We Investigate a multi-user imperfect CSI active RIS-assisted model taking into consideration a practical three-stage estimation error strategy and derive the SINR which reveals that the CSI error is coupled with power, beamforming and RIS elements.
    \item We derive three iterative algorithms to optimize the beamforming vectors at the base station (BS), the users' transmit power, and the gains and phase shifts of the RIS elements to solve the non-convex sum-rate maximization problem with a closed-form expression for the optimal beamforming vectors.
    \item Our simulations show that optimization is crucial for active RISs and that they are more sensitive to CSI errors than passive RISs.

\end{itemize}

\begin{figure}[t]
    \centering
    \includegraphics[width=0.85\columnwidth]{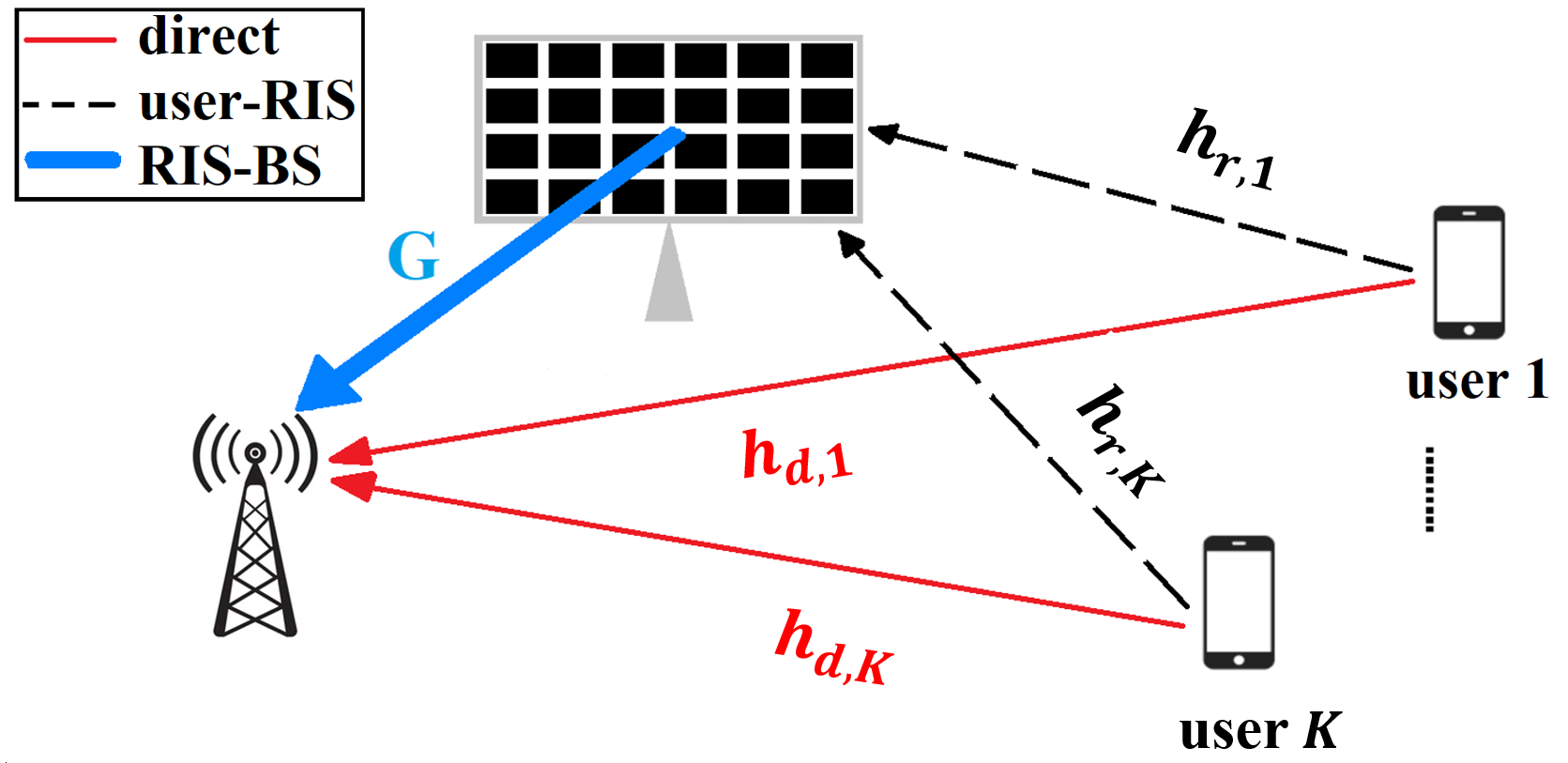}
    \caption{System Model}
    %\label{fig:my_label}
\end{figure}

% \emph{Notations}: $|\cdot|$ , $||\cdot||$ , $(\cdot)^*$ , $(\cdot)^T$ , $(\cdot)^H$ denote the absolute , square norm , the conjugate, transpose and conjugate transpose respectively. $j$ is the imaginary unit and $\operatorname{Re}\left\{ \cdot \right\}$, $\text{diag}(\cdot)$ , $\text{vec}(\cdot)$ denote the real, diagonalization and vectorization operators respectively. $\boldsymbol{0}_M$ and $\boldsymbol{I}_M$are the $M$-dimensional zero vector and identity matrix respectively.$\mathcal{CN}(0,\,\sigma^2)$ and $\mathcal{CCN}(0,\,\sigma^2)$ denote the complex normal and circular symmetric Gaussian distribution (CSGD) respectively with zero mean and variance $\sigma^2$.

\begin{figure*}[!t]
\normalsize
\begin{equation} \label{SINR}
\gamma_k=\frac{p_k\left|\boldsymbol{w}_k^H \boldsymbol{\hat{h}}_k\right|^2}{\underbrace{\sum^K_{i \neq k}{p_i\left|\boldsymbol{w}_k^H \boldsymbol{\hat{h}}_i\right|^2}}_{\text{interference}} +\underbrace{\underbrace{\sigma_n^2 \| \boldsymbol{w}_k^H \boldsymbol{\hat{G}\Phi} \|^2}_{\text{RIS noise}} +  \underbrace{\sum^K_{i=1}{p_i \left(  \sigma_{d,i}^2\|\boldsymbol{w}_k\|^2 + \sigma_{r,i}^2 \| \boldsymbol{\phi} \|^2 \| \boldsymbol{w}_k \|^2  \right)}+\sigma_G^2\sigma_n^2 \|\boldsymbol{w}_k \|^2 \| \boldsymbol{\phi} \|^2}_{\text{CSI error}}+ \underbrace{\sigma_0^2 \| \boldsymbol{w}_k \|^2}_{\text{receiver noise}}}_{\psi_k}}
\end{equation}
\hrulefill
\end{figure*}
%%%%%%%%%%%%%%%%%%%%%%%%%%%%%%%%%%%%%%%%%%%%
\section{ System Model and Problem Formulation}
An uplink scenario is considered as shown in Fig. 1 where $K$ single-antenna users transmit their signals to the BS equipped with $M$ antennas using a direct link and an RIS-assisted link. The RIS is a planar array with a total number of $N$ elements and each RIS element can introduce a phase shift and an amplification gain to the signal impinging on it. A controller with a separate link with the BS is attached to the RIS to acknowledge the CSI and decide the optimal REs. Denote $p_k$ as the transmit power of the $k$-th user, $S_k \in \mathbb{C}$ is its transmitted symbol, $\boldsymbol{w}_k \in {\mathbb{C} }^{M\times 1}$ is the beamforming vector applied by the BS to decode the $k$-th user's signal and $\boldsymbol{W}=[\boldsymbol{w}_1;\boldsymbol{w}_2 \cdots \boldsymbol{w}_K]$ is the beamforming matrix. $\boldsymbol{h}_{d,k} \in \mathbb{C}^{M\times 1}$ is the direct channel from the $k$-th user to the BS, $\boldsymbol{G}\in {\mathbb{C} }^{M\times N}$ is the channel from the RIS to the BS and ${\boldsymbol{h}}_{r,k}\in {\mathbb{C} }^{N\times 1}$ is the channel from the $k$-th user to the RIS. $\mathcal{K}=\left\{1,2,\cdots,K\right\}$ and $\mathcal{N}=\left\{1,2,\cdots,N\right\}$ denote the set of all users and all REs respectively. $\boldsymbol{a}=[a_1,a_2\cdots a_N]$ and $\boldsymbol{\theta}=[\theta_1,\theta_2\cdots \theta_N]$ denote the gain and phase shift vectors of the RIS respectively where $a_n \in [0,a_{max}] \forall n \in \mathcal{N}$ and ${\theta }_n \in [0,2\pi) \forall n \in \mathcal{N}$. The active REs vector is $\boldsymbol{\phi}=[a_1e^{j\theta_1},a_2e^{j\theta_2},\cdots,a_Ne^{j\theta_N}]$ and $\boldsymbol{\Phi}=\text{diag}(\boldsymbol{\phi})$. ${\boldsymbol{n}}_r\in {\mathbb{C} }^{N\times 1} \sim \mathcal{CN}(\boldsymbol{0}_N,\,\sigma_n^2\boldsymbol{I}_N)$ is the active REs noise and $\boldsymbol{n}_0 \in {\mathbb{C} }^{M\times 1} \sim \mathcal{CN}(\boldsymbol{0}_M,\,\sigma_0^2\boldsymbol{I}_M)$ is the receiver noise, both are modeled as complex Gaussian distributions.\\
\\
\hspace*{0.2cm} In practice, perfect channel estimation is difficult to attain. The estimation is done through at least two stages. First, the direct channel can be estimated by turning off the RIS in the absorbing mode while the reflected RIS channels can be estimated using the ON-OFF training reflection pattern technique \cite{9722893,9500469}. Second, the cascaded channels $\boldsymbol{H}_k=\boldsymbol{G}\text{diag}(\boldsymbol{h}_{r,k})$ are estimated instead of $\boldsymbol{G}$ and $\boldsymbol{h}_{r,k}$ separately. This step is the widely adopted for passive RISs. However, the active RIS noise can not be neglected like the passive one, which imposes knowing $\boldsymbol{G}$ since it is accompanied by the RIS noise. To address this issue, some works suggested deploying active sensors on the RIS with some RF chains \cite{9722893,9593172,9053976,9370097}. In \cite{9370097} compressive sensing and deep learning are used to estimate the channel $\boldsymbol{G}$ by sending pilots that are known to both the BS and the RIS. The hybrid channel estimation, i.e., estimating $\boldsymbol{G}$ using sensors and the cascaded $\boldsymbol{H}_k$ using the ON-OFF technique can reduce the training overhead compared to full separate channel estimation and also suits both time division duplexing and frequency division duplexing \cite{9722893}. This arises from the fact that, although $\boldsymbol{G}$ is generally high dimensional due to the multi-antenna BS, it is also quasi-static since the positions of BS and RIS are usually fixed \cite{9722893}. With imperfect CSI, the cascaded channel $\boldsymbol{H}_k$ is divided into two parts: the estimate and the error denoted by $\hat{\boldsymbol{H}_k}$ and $\triangle \boldsymbol{H}_k$, respectively. Similarly,  $\boldsymbol{G}=\boldsymbol{\hat{G}}+\triangle\boldsymbol{G}$ and the same applies for the direct channel $\boldsymbol{h}_{d,k}=\hat{\boldsymbol{h}}_{d,k}+\triangle \boldsymbol{h}_{d,k}$. Consider the combined channel of the $k$-th user as $\boldsymbol{h}_k=\boldsymbol{h}_{d,k}+ \boldsymbol{H}_{k}\boldsymbol{\phi}$ consequently, the combined channel is similarly divided into an estimate and error as $\hat{\boldsymbol{h}}_k=\hat{\boldsymbol{h}}_{d,k}+\hat{\boldsymbol{H}}_k\boldsymbol{\phi}$ and $\triangle \boldsymbol{h}_k=\triangle \boldsymbol{h}_{d,k}+\triangle \boldsymbol{H}_k \boldsymbol{\phi}$. The received signal at the BS on decoding the $k$-th user is
% \begin{multline}
%     y_k=\sqrt{p_k}\boldsymbol{w}_k^H \boldsymbol{\hat{h}}_k S_k + \sqrt{p_k}\boldsymbol{w}_k^H \boldsymbol{\triangle h}_k S_k + \sum^K_{i \neq k}{\sqrt{p_i}\boldsymbol{w}_k^H\boldsymbol{\hat{h}}_i S_i}+\\
%     \sum^K_{i \neq k}{\sqrt{p_i}\boldsymbol{w}_k^H \triangle \boldsymbol{h}_i S_i}  +\boldsymbol{w}_k^H \boldsymbol{\hat{G} \Phi n_r} + \boldsymbol{w}_k^H \triangle\boldsymbol{G \Phi n_r} +\boldsymbol{w}_k^H \boldsymbol{n}_0
% \end{multline}
\begin{multline}
    \hspace*{-0.25cm} y_k{=}\sqrt{p_k}\boldsymbol{w}_k^H \left( \boldsymbol{\hat{h}}_k + \boldsymbol{\triangle h}_k \right) S_k + \sum^K_{i \neq k}{\sqrt{p_i}\boldsymbol{w}_k^H \left( \boldsymbol{\hat{h}}_i + \triangle \boldsymbol{h}_i \right) S_i}\\
     +\boldsymbol{w}_k^H \left( \boldsymbol{\hat{G}} + \triangle\boldsymbol{G} \right) \boldsymbol{\Phi n_r} +\boldsymbol{w}_k^H \boldsymbol{n}_0
\end{multline}
% \begin{multline}
%     \hspace*{-0.25cm} y_k{=}\sqrt{p_k}\boldsymbol{w}_k^H \left( \hat{\boldsymbol{h}}_{d,k}+\triangle \boldsymbol{h}_{d,k}+\hat{\boldsymbol{H}}_k\boldsymbol{\phi} + \triangle \boldsymbol{H}_k \boldsymbol{\phi} \right) S_k + \\
%     \sum^K_{i \neq k}{\sqrt{p_i}\boldsymbol{w}_k^H \left( \hat{\boldsymbol{h}}_{d,i}+\triangle \boldsymbol{h}_{d,i}+\hat{\boldsymbol{H}}_i\boldsymbol{\phi} + \triangle \boldsymbol{H}_i \boldsymbol{\phi} \right) S_i}\\
%      +\boldsymbol{w}_k^H \left( \boldsymbol{\hat{G}} + \triangle\boldsymbol{G} \right) \boldsymbol{\Phi n_r} +\boldsymbol{w}_k^H \boldsymbol{n}_0
% \end{multline}
\noindent The error in estimation follows circular symmetric Gaussian distribution (CSGD) that is, $\triangle \boldsymbol{h}_{d,k}  \sim \mathcal{CN}(\boldsymbol{0}_M,\,\sigma_{d,k}^2\boldsymbol{I}_M)$ and $\text{vec}(\triangle \boldsymbol{H}_{k})  \sim \mathcal{CN}(\boldsymbol{0}_{MN},\,\sigma_{r,k}^2\boldsymbol{I}_{MN})$ \cite{9293148}, where $\text{vec}(\cdot)$ is the vectorization operator. Similarly, $\text{vec}(\triangle \boldsymbol{G})  \sim \mathcal{CN}(\boldsymbol{0}_{MN},\,\sigma_{G}^2\boldsymbol{I}_{MN})$ then the variance of $\boldsymbol{w}_k^H \boldsymbol{\triangle h}_i$ is given by $ \sigma_{d,i}^2||\boldsymbol{w}_k||^2 + \sigma_{r,i}^2 || \boldsymbol{\phi} ||^2 || \boldsymbol{w}_k ||^2$. Thus, the signal-to-interference plus noise ratio (SINR) of the $k$-th user received at the BS is expressed in (\ref{SINR}) at the top of this page. The sum-rate is given by 
\begin{equation} \label{sum_rate}
    R_s=\sum^K_{k=1}{\log_2{\left(1+\gamma_k\right)}}
\end{equation}
Then, the joint optimization problem is formulated as
\begin{equation*}
        \mathcal{P}1: \ \max_{\boldsymbol{W, \ P, \ \theta, \ a}}{\ \ \ R_s} \nonumber
\end{equation*}
\vspace{-0.5 cm}
\begin{gather}
 s.t. \ \ \ \ 0\leq p_k\leq p_{max}  \ \ \ \forall k \in \mathcal{K} \label{Cp} \\
    \| \boldsymbol{w}_k \|^2=1 \ \ \ \forall k \in \mathcal{K} \label{Cw}\\
    0\leq a_n\leq a_{max} \ \ \ \forall n \in \mathcal{N} \label{Ca} \\
    \theta_n \in [0,2\pi) \ \ \ \forall n \in \mathcal{N} \label{Cth}
\end{gather}
Constraints (\ref{Cp}), (\ref{Ca}) restrict the users' power, and active REs gains to their maximum $p_{max}$ and $a_{max}$, respectively, and ensures their non-negativity. Constraint (\ref{Cth}) preserves the range of phase shifts between $0$ and $2\pi$ and constraint (\ref{Cw}) restricts the norm of the beamforming vectors to unity. $\mathcal{P}1$ is a multi-dimensional optimization problem in $K \times M+K+2 \times N$ decision variables which is non-convex due the non-convexity of the objective function and the coupling of the four resources $\boldsymbol{W}$, $\boldsymbol{p}$, $\boldsymbol{\theta}$, $\boldsymbol{a}$ together in the objective function. To solve $\mathcal{P}1$, we first divide it into four sub-problems where in each sub-problem, one resource is optimized while the other are held fixed. Since each of the four sub-problems is still non-convex after this decoupling, we propose iterative approaches to overcome their non-convexity and solve them effectively. 
%%%%%%%%%%%%%%%%%%%%%%%
\section{Optimization of Beamforming, Power, Gain, and Phase Shifts}

\subsection{Power Optimization}
To find the optimal power while other resources are held fixed, we first use the fact that for two scalars $x$ and $y$ we have $\ln(1+x/y)=\ln(x+y)-\ln(y)$. Thus, using (\ref{SINR}), (\ref{sum_rate}) the objective function $R_s$ can be re-written as
\begin{multline} \label{R_T}
    R_T=\sum^K_{k=1}{\ln\left(\sum^K_{i=1}{p_i\left|\boldsymbol{w}_k^H \boldsymbol{\hat{h}}_i\right|^2}+\psi_k \right)}\\
    -\underbrace{\sum^K_{k=1}{\ln\left(\sum^K_{i \neq k}{p_i\left|\boldsymbol{w}_k^H \boldsymbol{h}_i\right|^2}+\psi_k\right)}}_{S_{ngv}}
\end{multline}
where $\psi_k$ is the term combining the RIS noise, the CSI error, and the receiver noise for the $k$-th user which is demonstrated in (\ref{SINR}). This transformation reduces the non-convexity of the objective function from sum of log of fractions to  sum of log of linear function in $\boldsymbol{p}$. To maintain a concave objective function on optimizing power, the negative sum $S_{ngv}$ in (\ref{R_T}) is transformed by using the first-order multi-variate Taylor expansion and neglecting the constant term since it does not affect the optimization. On considering a feasible expansion vector $\boldsymbol{p}^{(r)}$ at a certain iteration $(r)$, $S_{ngv}$ is transformed as
\begin{multline} \label{S_ngv}
S_{ngv}^{\ (r)}\approx \sum^K_{u=1} \left(p_u-p_u^{(r)}\right)\frac{\partial S_{ngv}}{\partial p_u}\left. \right|_{\boldsymbol{p}=\boldsymbol{p}^{(r)}} =\\
\sum^K_{u=1}\left(p_u-p_u^{(r)}\right)\left( 
 \frac{\sigma_{d,u}^2\|\boldsymbol{w}_u\|^2 + \sigma_{r,u}^2 \| \boldsymbol{\phi} \|^2 \| \boldsymbol{w}_u \|^2}{\sum^K_{i \neq u}{p_i^{(r)} \left| \boldsymbol{w}_u^H \boldsymbol{\hat{h}}_i \right|}+\psi_u^{(r)}} \right.\\
 +\left.  \sum^K_{k \neq u}{ \frac{ \left| \boldsymbol{w}_k^H \boldsymbol{\hat{h}}_u \right|^2 + \sigma_{d,u}^2\|\boldsymbol{w}_k\|^2 + \sigma_{r,u}^2 \| \boldsymbol{\phi} \|^2 \| \boldsymbol{w}_k \|^2 }{\sum^K_{i \neq k}{p_i^{(r)} \left| \boldsymbol{w}_k^H \boldsymbol{\hat{h}}_i \right|}+\psi_k^{(r)}} }    \right)
\end{multline}
 where $\psi_k^{(r)}=\psi_k|_{\boldsymbol{p}=\boldsymbol{p}^{(r)}}$ . With this transformation, the power optimization becomes convex and can be solved using the Lagrangian method expressed at the $r$-th iteration as
 \begin{multline}
     \mathcal{L}_P^{(r)}= \sum^K_{k=1}{\ln\left(\sum^K_{i=1}{p_i\left|\boldsymbol{w}_k^H \boldsymbol{\hat{h}}_i\right|^2}+\psi_k \right)} - S_{ngv}^{\ (r)} + \sum^K_{k=1}{\rho_kp_k} \\
     - \sum^K_{k=1}{\eta_k(p_k-p_{max})}
 \end{multline}
where $\boldsymbol{\rho }=[{\rho }_1,{\rho }_2,\dots ,{\rho }_K]$ and $\boldsymbol{\eta }=[{\eta }_1,{\eta }_2,\dots ,{\eta }_K]$ are the non-negative Lagrange multipliers for the power non-negativity and maximum power constraints respectively. By differentiating the Lagrangian function with respect to the $u$-th power we have
\begin{multline}
    \frac{\partial \mathcal{L}_P^{(r)}}{\partial p_u}= \sum^K_{k=1}{\frac{ \left| \boldsymbol{w}_k^H \boldsymbol{\hat{h}}_u \right|^2 + \sigma_{d,u}^2\|\boldsymbol{w}_k\|^2 + \sigma_{r,u}^2 \| \boldsymbol{\phi} \|^2 \| \boldsymbol{w}_k \|^2 }{\sum^K_{i=1}{p_i\left|\boldsymbol{w}_k^H \boldsymbol{\hat{h}}_i\right|^2}+\psi_k}}\\
    -  \frac{\sigma_{d,u}^2\|\boldsymbol{w}_u\|^2 + \sigma_{r,u}^2 \| \boldsymbol{\phi} \|^2 \| \boldsymbol{w}_u \|^2}{\sum^K_{i \neq u}{p_i^{(r)} \left| \boldsymbol{w}_u^H \boldsymbol{\hat{h}}_i \right|}+\psi_u^{(r)}}\\
   -\sum^K_{k \neq u}{ \frac{ \left| \boldsymbol{w}_k^H \boldsymbol{\hat{h}}_u \right|^2 + \sigma_{d,u}^2\|\boldsymbol{w}_k\|^2 + \sigma_{r,u}^2 \| \boldsymbol{\phi} \|^2 \| \boldsymbol{w}_k \|^2 }{\sum^K_{i \neq k}{p_i^{(r)} \left| \boldsymbol{w}_k^H \boldsymbol{\hat{h}}_i \right|}+\psi_k^{(r)}} } +\rho_u -\eta_u 
\end{multline}
The gradient descent algorithm (GDA) is used on the Lagrangian to obtain the optimal solution. Since the objective function is concave and the constraints are linear, strong duality holds and the Karush–Kuhn–Tucker conditions are sufficient, thus, the complementary slackness theorem holds \cite{boyd2004convex} and provides that at optimal $\boldsymbol{p}$, we have
\begin{equation}
    \rho_u p_u=\eta_u\left(p_u-p_{max}\right)=0 \ \ \forall u \in \mathcal{K}
\end{equation}
therefore, at optimal $\boldsymbol{p}$, either the Lagrange multipliers vanish or the optimal power is set to its lower or upper limits
\begin{equation} \label{opt_P}
p_u^{(r+1)}=\min\left(p_{max} \ , \ \max\left(0 \ , \ \chi_u^{(r)}\right)\right)
\end{equation}
where $\chi_u^{(r)}$ is obtained by applying the GDA with a step size $\nabla_P$ on the Lagrangian function when the Lagrangian multipliers are set to zero
\begin{equation}
\chi_u^{(r)}=p_u^{(r)}+\nabla_P \left.\frac{\partial {{\mathcal{L}}_P}^{\left(r\right)}}{\partial p_u}\right|_{\rho_u=0 , \ \eta_u=0  \ \forall u \in \mathcal{K}, \ \boldsymbol{p}=\boldsymbol{p}^{(r)} }
\end{equation}
To interpret (\ref{opt_P}), if $\chi_u^{(r)}$ exceeds $p_{max}$, the user transmits using full power. If $\chi_u^{(r)}$ is negative, the user's power is set to zero. If $\chi_u^{(r)} \in [0,p_{max}]$, then $\chi_u^{(r)}$ is indeed the optimal solution. The power optimization is provided in Algorithm \ref{alg_P}.
% After obtaining $S_{ngv}^{(r)}$ in a certain iteration $r$, the problem is solved to obtain the optimal power vector $\boldsymbol{p}^{(r+1)}$ which is used as the new expansion vector to obtain $S_{ngv}^{(r+1)}$ at the new iteration $r+1$. By re-approximating $S_{ngv}$ around the local optimal solution and solving the problem again, the solution approaches the exact. 
%%%%%%%%%%%%%%%%%%%%%%%%%%%%%%%%%%%%%%%%%%%%
\begin{algorithm}
\caption{Power Allocation}\label{alg_P}
\begin{algorithmic}
\State set $r=1$ , initialize $\boldsymbol{p}^{(1)}$
\While{$R_T^{(r+1)} - R_T^{(r)} > \epsilon$}
\State obtain $S_{ngv}^{(r)}$ around $\boldsymbol{p}^{(r)} $ using (\ref{S_ngv})
\State obtain the new optimal power $\boldsymbol{p}^{(r+1)} $ using (\ref{opt_P})
\State set $r=r+1$
\EndWhile
\end{algorithmic}
\end{algorithm}
%%%%%%%%
\begin{algorithm}
\caption{Gain and Phase shift optimization}\label{alg_t}
\begin{algorithmic}
\State set $r=1$ , initialize $\boldsymbol{\theta}^{(1)}$
\While{$R_T^{(r+1)} - R_T^{(r)} > \epsilon$}
\State obtain the optimal phase shifts using (\ref{opt_t})
\State set $r=r+1$
\EndWhile

\State set $r=1$ , initialize $\boldsymbol{a}^{(1)}$
\While{$R_T^{(r+1)} - R_T^{(r)} > \epsilon$}
\State obtain the optimal gains using (\ref{opt_a})
\State set $r=r+1$
\EndWhile
\end{algorithmic}
\end{algorithm}
%%%%%%%%
\begin{algorithm}
\caption{Optimal beamforming}\label{alg_W}
\begin{algorithmic}
\State set $r=1$ , initialize $\boldsymbol{w}_k \ \forall k \in \mathcal{K}$, set scaling factor $\delta=0.9$
\While{$R_W^{(r+1)} - R_W^{(r)} > \epsilon$}
\State For all $K$ users
\State set $\mu_k=1$
\State set $\alpha_k=\gamma_k$
\State calculate $\beta_k$ using (\ref{beta_k})
\While {$| \ ||\boldsymbol{w}_k^{opt}||^2-1 | > \epsilon$}
\State obtain $\boldsymbol{w}_k^{opt}$ using (\ref{opt_W})
\If{$||\boldsymbol{w}_k^{opt}||<1$} 
    \State $\mu=\delta\mu$
\Else
    \State $\mu=\mu/\delta$
\EndIf 
\EndWhile
\State set $r=r+1$
\EndWhile
\end{algorithmic}
\end{algorithm}
%%%%%%%%%%%%%%%%%%%%%%%%%%%%%%%%%%%%%%%%%%%%
\begin{figure*}
    \begin{equation*} 
    R_W=\sum^K_{k=1}{\left(\ln(1+\alpha_k)-\alpha_k\right)}+\sum^K_{k=1}{2\sqrt{(1+\alpha_k) p_k}\operatorname{Re}\left\{ \beta_k^* \boldsymbol{w}_k^H\boldsymbol{\hat{h}}_k \right\}}   
\end{equation*}
\begin{equation} \label{CFQT}
    -\sum^K_{k=1}{|\beta_k|^2 \left[\sum^K_{i=1}{p_i\left|\boldsymbol{w}_k^H \boldsymbol{\hat{h}}_i\right|^2} {+}\sigma_n^2 \| \boldsymbol{w}_k^H \boldsymbol{\hat{G}\Phi} \|^2 {+}  \sum^K_{i=1}{p_i \left(  \sigma_{d,i}^2\|\boldsymbol{w}_k\|^2 {+} \sigma_{r,i}^2 \| \boldsymbol{\phi} \|^2 \| \boldsymbol{w}_k \|^2  \right)}+\sigma_G^2\sigma_n^2 \|\boldsymbol{w}_k \|^2 \| \boldsymbol{\phi} \|^2 {+} \sigma_0^2 \| \boldsymbol{w}_k \|^2 \right]}
\end{equation}

\hrulefill
\begin{equation} \label{opt_W}
    \boldsymbol{w}_k^{opt}{=} \sqrt{(1+\alpha_k)p_k} \beta_k^* \left[  \mu_k\boldsymbol{I}_M {+}   |\beta_k|^2 \left[   \sum^K_{i=1}{p_i \boldsymbol{\hat{h}}_i \boldsymbol{\hat{h}}_i^H} + \sigma_n^2 \boldsymbol{\hat{G}} \boldsymbol{\Phi} \boldsymbol{\Phi}^H \boldsymbol{\hat{G}}^H {+} \left[ \sigma_G^2\sigma_n^2  \| \boldsymbol{\phi} \|^2  {+} \sigma_0^2 {+} \sum^K_{i=1}{p_i \left(  \sigma_{d,i}^2 {+} \sigma_{r,i}^2 \| \boldsymbol{\phi} \|^2  \right)  }  \right] \boldsymbol{I}_M  \right]  \right]^{-1} \hspace{-0.2cm}\boldsymbol{\hat{h}}_k
\end{equation}

\hrulefill

%\vspace*{4pt}
\end{figure*}
%%%%%%%%%%%%%%%%%%%%%%%%%%%%%%%%%%%%%%%%%
\subsection{Gain Optimization}
To find the optimal gain of the REs, the GDA is used. Since the objective function is non-convex, the obtained solution is a stationary local optimal solution that is not necessarily the global optimal. To abide by constraint (\ref{Ca}), the projected GDA is used and the gain at iteration $r+1$ is given by
\begin{equation} \label{opt_a}
\boldsymbol{a}^{(r+1)}=Q_{a}\left(\boldsymbol{a}^{(r)}+ \nabla_{a} \left.\frac{\partial R_s}{\partial \boldsymbol{a}}\right|_{\boldsymbol{a}=\boldsymbol{a}^{(r)}} \right)
\end{equation}
where $\nabla_{a}$ is the step size for the GDA and $Q_{a}(\cdot)$ is the projection function given by \cite{boyd2004convex}
\begin{equation} \label{Q_a}
    Q_{a}( \boldsymbol{a}^{(r)} )=\min_{\boldsymbol{a} \in [0,a_{max}]}{  \| \boldsymbol{a}-\boldsymbol{a}^{(r)} \| }
\end{equation}
In the projected GDA, the projection function (\ref{Q_a}) aims to project any solution outside the feasible interval $[0,a_{max}]$ inside it. It is noted that (\ref{Q_a}) is itself an optimization problem to minimize the difference between the unbounded solution and the solution in the feasible set. It is easy to see that, for the feasible set $[0,a_{max}]$, the closest projection of any negative value is zero and the closet projection of any value exceeding $a_{max}$ is $a_{max}$ itself, therefore
\begin{equation}
    Q_a(a)=\min\left( \max(0,a) , a_{max} \right)
\end{equation}
%%%%%%%%%%%%%%%%%%%%%%%%%%%%%%%%%%%%
\subsection{Phase Shifts Optimization}
To find the optimal phase shifts, the projected GDA is used similar to the gain optimization, which results in
\begin{equation} \label{opt_t}
\boldsymbol{\theta}^{(r+1)}=Q_{\theta}\left(\boldsymbol{\theta}^{(r)}+ \nabla_{\theta} \left.\frac{\partial R_s}{\partial \boldsymbol{\theta}}\right|_{\boldsymbol{\theta}=\boldsymbol{\theta}^{(r)}} \right)
\end{equation}
where $\nabla_{\theta}$ is the step size for the GDA and $Q_{\theta}(\cdot)$ is the projecting function that projects any solution outside the intended interval $[0,2\pi)$ inside it. The projection can be expressed as
 \begin{equation} \label{q_t}
     Q_{\theta}(\theta)=\arcsin(\sin(\theta))
 \end{equation}
  Eq. (\ref{q_t}) uses the fact that for any unbounded phase shift $\theta_{unb}$, $\sin{(\theta_{unb})}$ produces real values between $-1$ and $1$ even if $\theta_{unb}$ is outside $[0,2\pi)$. By applying the inverse sine, the obtained bounded solution $\theta_b=\arcsin(\sin(\theta_{unb}))$ will be in the intended interval $[0,2\pi)$. Both $\theta_{unb}$ and $\theta_b$ are equivalent in the complex exponential function except that $\theta_b$ is in the feasible set.

\begin{figure*}
        \includegraphics[width=18cm]{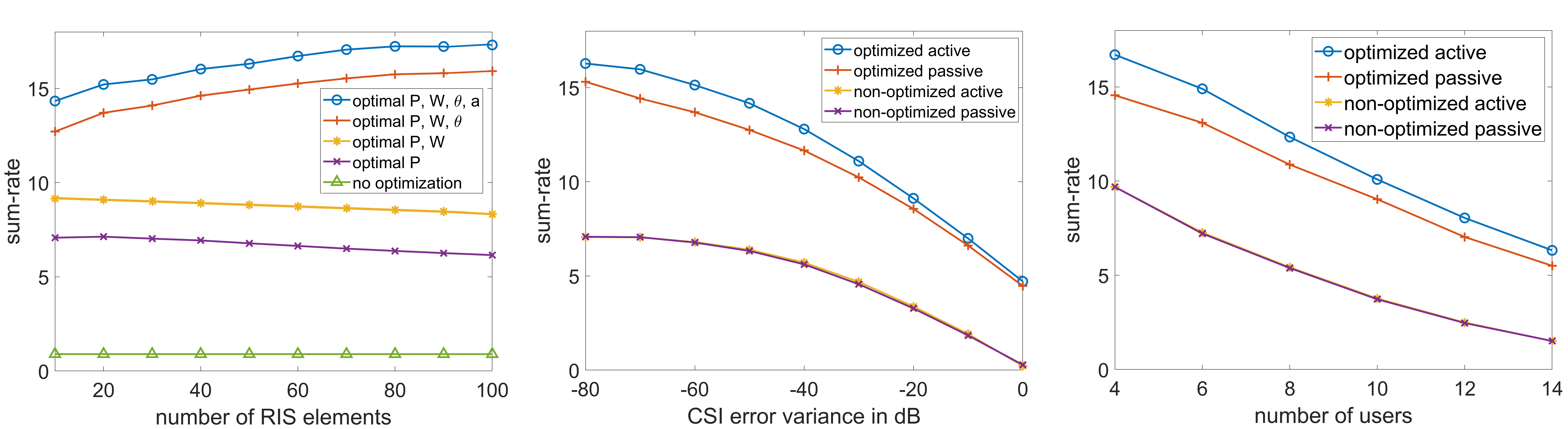}
        \caption{Sum-rate vs $N$ \hspace{2.6cm} Fig. 3: Sum-rate vs CSI error variance \hspace{0.2cm} Fig. 4: Sum-rate vs $K$}
        \label{fig:my_label}
\end{figure*}

\subsection{Beamforming Vectors Optimization}
The fractional programming (FP) approach is used to extract the beamfroming vectors out of the logarithmic function. By using the closed-form approach and applying the quadratic transform in \cite{8310563}, the objective function $R_s$ is transformed to the form provided in (\ref{CFQT}) at the top of this page where $ \forall k \in \mathcal{K} \ \alpha_k$ and $\beta_k$ are the auxiliary variables introduced by the quadratic transform and are updated in each iterations as \cite{8310563}
\begin{equation}
        \alpha_k=\gamma_k 
\end{equation}
\begin{equation} \label{beta_k}
    \beta_k=\frac{ \sqrt{(1+\alpha_k) p_k}  \boldsymbol{w}_k^H\boldsymbol{\hat{h}}_k }{\sum^K_{i=1}{p_i\left|\boldsymbol{w}_k^H \boldsymbol{\hat{h}}_i\right|^2} + \psi_k}
\end{equation}
The Lagrangian function is
\begin{equation} \label{L_W}
    \mathcal{L}_W=R_W-\sum^K_{k=1}{\mu_k\left(\| \boldsymbol{w}_k \|^2-1\right)}
\end{equation}

\noindent where $\mu_k$ is the Lagrange multiplier introduced to satisfy the unity norm constraint (\ref{Cw}) for the $k$-th user. By re-writing $|\boldsymbol{w}_k^H \boldsymbol{\hat{h}}_i|^2$ as $\boldsymbol{w}_k^H\boldsymbol{\hat{h}}_i \boldsymbol{\hat{h}}_i^H \boldsymbol{w}_k$ and $\| \boldsymbol{w}_k^H \boldsymbol{\hat{G}\Phi} \|^2$ as $\boldsymbol{w}_k^H\boldsymbol{\hat{G}} \boldsymbol{\Phi} \boldsymbol{\Phi}^H \boldsymbol{\hat{G}}^H\boldsymbol{w}_k$ and differentiating the Lagrangian function (\ref{L_W}) with respect to $\boldsymbol{w}_k$ and equating it to zeros, we obtain the optimal solution given in (\ref{opt_W}). The beamforming optimization process is summarized in Algorithm \ref{alg_W}.

\subsection{Convergence and Complexity Analysis}
All our algorithms run till the value of the sum-rate in two successive iterations falls below a certain threshold $\epsilon$. The convergence of FP is studied in \cite{8310563} and has been used in several works \cite{9568854},\cite{8982186}. For power optimization, any feasible expansion vector $\boldsymbol{p}^{(r)}$ makes the optimization problem convex. Thus, Algorithm \ref{alg_P} converges to the optimal solution since any local optimum is globally optimal in convex problems. For the phase shifts and gain optimization, the GDA can easily converge to a locally optimal solution as stated before with feasible initialization and appropriate step size. The complexity of power, gain, phase shift optimization are $\mathcal{O}(L_p \times K)$, $\mathcal{O}(L_a \times N)$, $\mathcal{O}(L_{\theta} \times N)$, respectively, where $L$ denotes number of loops taken by the GDA to converge for the relevant resource. In general, the number of loops is dependent on the step size and error tolerance $\epsilon$. In our simulations, for $\epsilon=10^{-2},\nabla_P=0.1$, and $N=60$, the average of $L_P$ was 3, 8 at $K=5$, $K=10$, respectively. For $\epsilon= 10^{-2},\nabla_{\theta}=0.1$, and $K=5$, the average of $L_{\theta}$ was 25, 37 at $N=30$, $N=80$, respectively. For $\epsilon=0.5 \times 10^{-2},\nabla_a=0.1$, and $K=5$, the average of $L_a$ was 50, 80 at $N=30$, $N=80$, respectively. Compared to \cite{9568854}, the complexity of GDA is smaller than the joint use of FP and alternating direction method of multipliers.
%%%%%%%%%%%%%%%%%%%%%%%%%%%%%%%%%%%%%%%%%%%%%

\section{Simulation results}%%%%%%%%%%%%%%%%%%%%%%%%%%%
 We consider all channels follow a Rayleigh distribution with normalized power and normalize the sum-rate to unit bandwidth. We assume $M{=}10$, $p_{max}{=}20$ dBm, $a_{max}{=}4$, $\nabla_P{=}\nabla_{\theta}{=}\nabla_a{=}0.1$, $\sigma_{n}^2{=}-60$ dB, and $\forall i \in \mathcal{K} \ \sigma_0^2{=}\sigma^2_{d,i}{=}\sigma_{r,i}^2{=}-70$ dB. All resources are initialized with random values in their feasible set. Any unoptimized resources are kept random.\\
\hspace*{0.5cm} Fig. 2 shows the sum-rate vs the number of RIS elements at $K=5$. In the case of no optimization, the sum-rate is around 1 b/sec/Hz. Power optimization can enhance the sum-rate to above 6 b/sec/Hz, and the joint power and beamforming optimization improves it to above 8 b/sec/Hz. Deploying a passive RIS increases the sum-rate by 75 \% at $N=30$ and by 95 \% at $N=100$ compared to the joint power and beamforming optimization. On the other hand, active RIS improve it by 90\% at $N=30$ and by 110 \% at $N=100$.\\
\hspace*{0.5cm} Fig. 3 shows the sum-rate vs the CSI error variance $\forall \ i \in \mathcal{K} \ \sigma_{d,i}^2,\sigma_{r,i}^2, \ \sigma_G^2$ at $K=5$, and $N=80$ and all CSI errors are assumed to be equal. Four setups are plotted: a) optimized active RIS, where all of the four resources are optimized, b) optimized passive RIS, where $\boldsymbol{p,W,\theta}$ are optimized, c) non-optimized active RIS, where only $\boldsymbol{p,W}$ are optimized, and d) non-optimized passive RIS, where only $\boldsymbol{p,W}$ are optimized. The non-optimized active and non-optimized passive curves nearly coincide on each other. In fact, active RIS improve the SINR by providing a gain but at the same time amplifies the noise which proves the importance of gain optimization. At high error variances, the performance of active RISs is deteriorated more than passive RIS since an active RIS is affected by the estimation error $\sigma_G^2$ which concludes that active RISs are more affected by CSI errors compared to passive ones.\\
\hspace*{0.5cm} Fig. 4 shows the sum-rate vs the number of users at $N=80$ for the same four setups mentioned in Fig. 3. The sum-rate is deteriorated by increasing $K$ due to the worst case considered of double Rayleigh fading with imperfect CSI and the RIS noise. Optimizing the power and beamforming vectors can support up to 12 users with a sum-rate below 2 b/sec/Hz. Optimized passive RISs improve it to 5.5 b/sec/Hz, while active RIS increases it to above 6 b/sec/Hz.
\section{Conclusion}
The problem of optimally allocating power, active RE gains, and phase shifts along with the beamforming vectors at the base station for an active RIS-assisted communications system is addressed with imperfect CSI to maximize the sum-rate. Simulation results show that the superiority of active RIS at the expense of higher sensitivity to CSI imperfections at high error variances.
\bibliographystyle{IEEEtran} 
\bibliography{REF}

\end{document}